\documentclass[twocolumn]{autart}    

\usepackage{graphicx}          
\usepackage{graphicx} 
\usepackage[fleqn]{amsmath} 
\usepackage{amssymb}  
\usepackage{color}
\usepackage{multicol}
\usepackage{epstopdf}

\usepackage[colorinlistoftodos,textsize=small,textwidth=2cm]{todonotes}

\begin{document}

\begin{frontmatter}

\title{A port-Hamiltonian approach to the control of nonholonomic systems} 


\author[1st]{Joel Ferguson}\ead{Joel.Ferguson@uon.edu.au},    
\author[2nd]{Alejandro Donaire}\ead{alejandro.donaire@qut.edu.au},               
\author[3rd]{Christopher Renton}\ead{Christopher.Renton@newcastle.edu.au},  
\author[1st]{Richard H. Middleton}\ead{Richard.Middleton@newcastle.edu.au}

\address[1st]{School of Electrical Engineering and Computing and PRC CDSC, The University of Newcastle, Callaghan, NSW 2308, Australia.}  
\address[2nd]{Department of Electrical Engineering and Information Theory and PRISMA Lab, University of Naples Federico II, Napoli 80125, Italy, and with the School of Electrical Eng. and Comp. Sc. of the Queensland University of Technology, Brisbane, QLD, Australia. }             
\address[3rd]{School of Engineering, The University of Newcastle, Callaghan, NSW 2308, Australia.}        


\begin{abstract}                          
In this paper a method of controlling nonholonomic systems within the port-Hamiltonian (pH) framework is presented. It is well known that nonholonomic systems can be represented as pH systems without Lagrange multipliers by considering a reduced momentum space. Here, we revisit the modelling of these systems for the purpose of identifying the role that physical damping plays. Using this representation, a geometric structure generalising the well known chained form is identified as \textit{chained structure}. A discontinuous control law is then proposed for pH systems with chained structure such that the configuration of the system asymptotically approaches the origin. The proposed control law is robust against the damping and inertial of the open-loop system. The results are then demonstrated numerically on a car-like vehicle.
\end{abstract}

\end{frontmatter}

\section{Introduction}

The control problem of set-point regulation for nonholonomic systems has been widely studied within the literature \cite{Astolfi1996,Bloch2003,Kolmanovsky,Tian2002,Murray1991a}. This problem is inherently difficult as nonholonomic systems do not satisfy Brockett's necessary condition for smooth stabilisation which implies that they cannot be stabilised using smooth control laws, or even continuous control laws \cite{Bloch2003}. In response to these limitations, the control community has utilised several alternate classes of controllers to stabilise nonholonomic systems including time-varying control \cite{Samson1995,Tian2002}, switching control \cite{Lee2007} and discontinuous control \cite{Astolfi1996,Fujimoto2012}.

One approach that has been utilised to solve the control problem is to assume that the system has a particular kinematic structure known as \textit{chained form} \cite{Murray1991a}. This structure was previously utilised in \cite{Astolfi1996} to propose a discontinuous control law to achieve set-point regulation for this class of systems. While it may seem restrictive to assume this kinematic structure, many systems of practical importance have been shown to be of chained form under suitable coordinate and input transformations. Examples of this are the kinematic car \cite{Murray1991a} and the $n$-trailer system \cite{Sordalen1993}. This form of kinematic structure plays a central role in the developments presented here.

Dynamic models for many nonholonomic systems (ie. systems with drift) are able to be formulated within the port-Hamiltonian framework where the constraints enter the dynamics equations as Lagrange multipliers \cite{goldstein1965classical}. It was shown in \cite{VanDerSchaft1994} that the Lagrange multipliers, arising from the constraint equations, can be eliminated from the pH representation of such systems by appropriately reducing the dimension fo the momentum space. Interestingly, the reduced equations have a non-canonical structure and the dimension of the momentum space is less than the dimension of the configuration space. It was further shown in \cite{Maschke1994} that stabilisation of the pH system can easily be achieved using the reduced representation via potential energy shaping. Asymptotic stability, however, was not considered in that work.

While the control of nonholonomic systems has been extensively studied within the literature, control methods that exploit the natural passivity of these systems are rather limited. Some exceptions to this trend are the works \cite{Donaire2015,Delgado2016,Muralidharan2009} which all utilised smooth control laws to achieve some control objective. In each of these cases, the control objective was to stabilise some non-trivial submanifold of the configuration space with characteristics such that Brockett's condition does not apply. Similar to this approach, a switching control law for a 3-degree of freedom mobile robot was proposed in \cite{Lee2007}. Each of the individual control laws used in the switching scheme were smooth and stabilised a sub-manifold of the configuration space. The stabilised sub-manifolds were chosen such that their intersection was the origin of the configuration space. Using a switching heuristic, the switching control law was able to drive the 3-degree of freedom robot to a compact set containing the origin. 

In our previous work \cite{Ferguson2016}, we considered a switching control law for the Chaplygin sleigh where each of the individual control laws were potential energy-shaping controllers were the target potential energy was a discontinuous function of the state. Each of the controllers stabilised a submanifold of the configuration space where the stabilised sub-manifolds were chosen such that they intersect at the origin. A switching heuristic was then proposed such that the system converged to the origin of the configuration space asymptotically. Likewise, asymptotic stability of 3-degree of freedom nonholonomic systems was considered in \cite{Fujimoto1999,Fujimoto2012} where the proposed approach was to use a potential energy-shaping control law where the target potential energy was a discontinuous function of the state. This approach has the advantage of not requiring any switching heuristic to achieve convergence.

In this paper, inspired by the works \cite{Astolfi1996} and \cite{Fujimoto2012}, we propose a discontinuous potential energy-shaping control law for a class of nonholonomic systems\footnote{A short version of this paper has been accepted for presentation at LHMNC 2018 \cite{Ferguson2018}. The conference version considers control of the Chaplygin sleigh system using a simplified version of the control law presented here. Lemma \ref{FuncInequality}, Proposition \ref{z1neq0} and Proposition \ref{w1Zero} can be found within the conference version. The extension to $n$-dimensional systems, the presented example and all other technical developments are original to this work.}. First, the procedure to eliminate the Lagrange multipliers from the pH representation of a nonholonomic system proposed in \cite{VanDerSchaft1994} is revisited and the role of physical damping is defined. Then, considering the reduced representation of the system, a special geometric structure that generalises the well know chained form is identified and called \textit{chained structure}. A discontinuous control law, with the interpretation of potential energy shaping together with damping injection, is then proposed for $n$-degree of freedom pH systems with chained structure such that the configuration asymptotically converges to the origin. The controller is shown to be robust against the damping and inertial properties of the open-loop system.

\textbf{Notation:}
Given a scalar function $f(x):\mathbb{R}^n\to \mathbb{R}$, $\nabla_x f$ denotes the column of partial derivatives $\begin{bmatrix} \frac{\partial}{\partial x_1}f & \cdots & \frac{\partial}{\partial x_n}f \end{bmatrix}^\top$. For a vector valued function $g(x)\in\mathbb{R}^m$, $\frac{\partial}{\partial x} g$ denotes the standard Jacobian matrix. $I_n$ and $0_n$ denote the $n\times n$ identity and zero matrices, respectively. $0_{n\times m}$ is the $n\times m$ zero matrix.

\section{Problem formulation}\label{prob}
This work is concerned with mechanical systems that are subject to constraints that are non-integrable, linear combination of generalised velocities:
\begin{equation}\label{PHS:PfaffianFactor}
	G_c^\top( q)\dot{ q} = 0_{k\times 1},
\end{equation}
where $q\in\mathbb{R}^n$ is the configuration, $k$ is the number of linearly independent constraints and $G_c \in \mathbb{R}^{n\times k}$ is full rank.
Such constraints are called nonholonomic, Pfaffian constraints \cite{Choset2005} and naturally arises when considering non-slip conditions of wheels \cite{Bloch2003}. For the remainder of the paper, the term nonholonomic is used to refer to constraints of the form \eqref{PHS:PfaffianFactor}.

Nonholonomic constraints do not place a restriction on achievable configurations of the system, but rather, restricts the valid paths of the system through the configuration space. Mechanical systems with nonholonomic constraints can be modelled as pH systems where the constants appear as Lagrange multipliers \cite{Maschke1994}:
\begin{equation}\label{PHS:canonicalMomentumPfaffian}
	\begin{split}
		\begin{bmatrix}
			\dot{q} \\
			\dot{p}_0
		\end{bmatrix}
		&=
		\begin{bmatrix}
			0_n & I_n \\
			-I_n & -D_0
		\end{bmatrix}
		\begin{bmatrix}
			\nabla_{q}\mathcal{H}_0 \\
			\nabla_{p_0}\mathcal{H}_0
		\end{bmatrix}
		+
		\begin{bmatrix}
			0_{n\times m} & 0_{n\times k} \\
			G_0 & G_c \\
		\end{bmatrix}
		\begin{bmatrix}
			u \\
			\lambda
		\end{bmatrix} \\
		y
		&=
		G_0^\top
		\nabla_{p_0}\mathcal{H}_0
		+
		G_c^\top
		\nabla_{p_0}\mathcal{H}_0
		=
		G_0^\top
		\nabla_{p_0}\mathcal{H}_0 \\
		\mathcal{H}_0(q,p)
		&=
		\underbrace{
		\frac12 p_0^\top M_0^{-1}(q) p_0
		}_{\mathcal{T}_0}
		+
		\mathcal{V}(q),
	\end{split}
\end{equation}
where $p_0 \in \mathbb{R}^n$ is the momentum, $u,y\in\mathbb{R}^m$, with $m = n-k$, are the input and output respectively, $G_0(q)\in\mathbb{R}^{n\times m}$ is the input mapping matrix, $\lambda\in\mathbb{R}^k$ are the Lagrange multipliers corresponding to the constraints \eqref{PHS:PfaffianFactor}, $D_0(p_0, q) = D_0(p_0, q)^\top \geq 0$ contains physical damping terms, $M_0(q) = M_0^\top(q) > 0$ is the mass matrix, $\mathcal{T}_0$ is the kinetic energy and $\mathcal{V}( q) > 0$ is the potential energy \cite{Gomez-Estern2004}, \cite{Ortega2004}. It is assumed that the matrix $\begin{bmatrix} G_0(q) & G_c(q) \end{bmatrix}\in \mathbb{R}^{n\times n}$ is full rank. The constraint equation \eqref{PHS:PfaffianFactor} has been used to determine $G_c^\top\nabla_{p_0}\mathcal{H}_0 = G_c^\top\dot{q} = 0_{m\times 1}$ to simplify the output equation. \\

\textbf{Problem statement:}
	Given the nonholonomic pH system \eqref{PHS:canonicalMomentumPfaffian}, design a discontinuous control law $u=u(q,p_0)$ such that $\lim_{t\to\infty}q(t) = 0_{n\times 1}$.\\

\begin{figure}
	\centering
	\includegraphics[trim = 0mm 3mm 0mm 0mm, clip, width=1.0\linewidth]{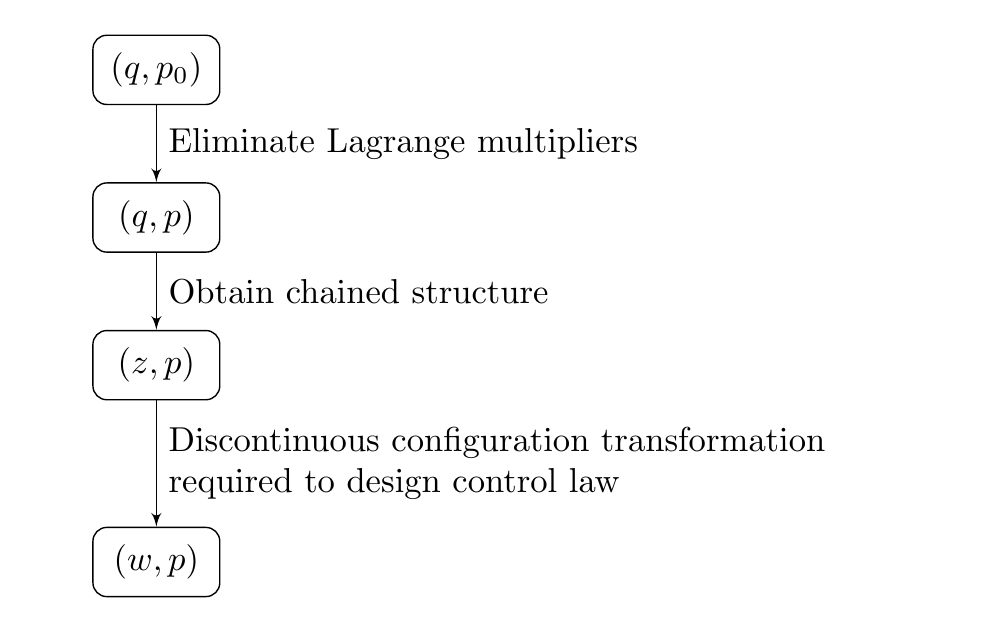}
	\caption{The progression of coordinate transformations and their respective purposes.}
	\label{CoorTrans}
\end{figure}

Throughout this paper, several coordinate transformations will be performed on the nonholonomic system \eqref{PHS:canonicalMomentumPfaffian} in order to address the problem statement. Figure \ref{CoorTrans} summarises the coordinate transformations utilised and states their respective purposes.

\section{Elimination of Lagrange multipliers}\label{PHS}

In this section, the system \eqref{PHS:canonicalMomentumPfaffian} is simplified by eliminating the Lagrange multipliers from the pH representation. As was done in \cite{VanDerSchaft1994}, this simplification is achieved via a reduction of the dimension of the momentum space. The presented formulation explicitly considers the role of physical damping which will be utilised in the following sections. To this end, we recall the following lemma:
\begin{lem}[Section 3, \cite{Viola2007}]\label{prop:momentumtransformation}
	Let $Q_0(q)\in\mathbb{R}^{n\times n}$ be any invertible matrix and define $\tilde p = Q_0^\top(q) p_0$. Then, the dynamics \eqref{PHS:canonicalMomentumPfaffian} can be equivalently expressed in coordinates $(q,\tilde p)$ as:
	\begin{equation}\label{PHS:TransformedMomentumPfaffian}
		\begin{split}
			\begin{bmatrix}
				\dot{ q} \\
				\dot{\tilde{ p}}
			\end{bmatrix}
			&=
			\begin{bmatrix}
				0_n & Q_0 \\
				-Q_0^\top & \tilde{C}-\tilde{D}
			\end{bmatrix}
			\begin{bmatrix}
				\nabla_{q}\mathcal{H} \\
				\nabla_{\tilde{p}}\mathcal{H} \\
			\end{bmatrix}
			+
			\begin{bmatrix}
				0_{n\times m} & 0_{n\times k} \\
				G & \tilde{G} \\
			\end{bmatrix}
			\begin{bmatrix}
				u \\
				\lambda \\
			\end{bmatrix} \\
			y 
			&= 
			\begin{bmatrix}
				G^\top \\
				\tilde{G}^\top
			\end{bmatrix}
			\nabla_{\tilde{p}}\mathcal{H} \\
			\tilde{\mathcal{H}}(q,\tilde{ p}) 
			&\triangleq 
			\mathcal{H}_0(q,Q_0^{-\top}\tilde{ p}) 
			= 
			\underbrace{
			\frac12 \tilde{ p}^\top \tilde{M}^{-1}(q) \tilde{ p} 
			}_{\tilde{\mathcal{T}}}
			+ 
			\mathcal{V}(q),
		\end{split}
	\end{equation}
	where
	\begin{equation}\label{PHS:TransformedMomentumPfaffianDefinitions}
		\begin{split}
			\tilde{D}(q,\tilde{ p}) 
			&= 
			Q_0^\top( q) D_0(Q_0^{-\top}( q) \tilde{ p}, q) Q_0( q) \\
			\tilde{C}(q,\tilde{ p}) 
			&= 
			Q_0^\top( q) \left\{ \frac{\partial^\top}{\partial q}\left[Q_0^{-\top}( q) \tilde{ p}\right] \right.\\
			&\left.\phantom{------}- 
			\frac{\partial}{\partial q}\left[Q_0^{-\top}( q) \tilde{ p}\right] \right\} Q_0( q) \\
			\tilde{M}( q) 
			&= 
			Q_0^\top( q) M_0( q) Q_0( q) \\
			G( q) 
			&= 
			Q_0^\top( q) G_0( q) \\
			\tilde{G}( q)
			&= 
			Q_0^\top( q) G_c( q).
		\end{split}
	\end{equation}
\end{lem}

It is now shown that for an appropriate choice of $Q_0$, \eqref{PHS:TransformedMomentumPfaffian} can be equivalently described in a reduced momentum space without Lagrange multipliers. To see this, let $G_c^\perp( q)$ be a left annihilator of $G_c( q)$ with rank $m = n-k$. Using this definition we propose the new momentum variable
	\begin{equation}\label{PHS:momentumTransformationPartition}
		 \tilde{ p} =
		 \begin{bmatrix}
		     \mu \\
		      p
		 \end{bmatrix}
		 =
		 \underbrace{
		 \begin{bmatrix}
		     A^\top( q) \\
		     Q^\top( q)
		 \end{bmatrix}
		 }_{Q_0^\top( q)}
		  p_0,
	\end{equation}
for the system \eqref{PHS:canonicalMomentumPfaffian} where $\mu \in \mathbb{R}^k$, $ p \in \mathbb{R}^m$ and $Q = (G_c^\perp)^\top$.
\begin{lem}\label{Lemma:QInvert}
    Consider the matrix $Q_0$ in \eqref{PHS:momentumTransformationPartition}. If $G_c^\top A$ is invertible, then $Q_0$ is also invertible.
\end{lem}
\begin{pf}
	The proof is provided in the Appendix.
\end{pf}
The matrix $A$ is then chosen such that $G_c^\top A$ is invertible which implies that $Q_0$ is invertible by Lemma \ref{Lemma:QInvert}.

\begin{prop}\label{PHS:nonholreduc}
	Consider system \eqref{PHS:canonicalMomentumPfaffian} under the change of momentum \eqref{PHS:momentumTransformationPartition}. The system dynamics can equivalently expressed as
	\begin{equation}\label{PHS:ReducMomentum}
		\begin{split}
			\begin{bmatrix}
			\dot{ q} \\
			\dot{ p} \\
			\end{bmatrix}
			&=
			\begin{bmatrix}
			0_n & Q \\
			-Q^\top & C-D
			\end{bmatrix}
			\begin{bmatrix}
			\nabla_q\mathcal{H} \\
			\nabla_p\mathcal{H}
			\end{bmatrix} 
			+
			\begin{bmatrix}
			0_{n\times m} \\
			G \\
			\end{bmatrix}
			u \\
			y 
			&= 
			G^\top \frac{\partial \mathcal{H}}{\partial  p} \\
			\mathcal{H}( p, q) 
			&= 
			\underbrace{
			\frac12  p^\top M^{-1}  p 
			}_{\mathcal{T}}
			+ 
			\mathcal{V},
		\end{split}
	\end{equation}
	where
	\begin{equation}\label{PHS:ReducMomentumMatricies}
	    \begin{split}
	        G( q) 
	        &= 
	        Q^\top( q) G_0( q) \\
	        D( p, q) 
	        &= 
	        Q^\top( q) D_0( q, p_0) Q( q) \\
	        C( p, q) 
	        &= 
	        Q^\top( q) \left\{ \frac{\partial^\top}{\partial q}\left[ M_0 Q \left( Q^\top M_0 Q \right)^{-1}  p\right] \right.\\
	        &\left.\phantom{---} - \frac{\partial}{\partial q} \left[ M_0 Q \left( Q^\top M_0 Q \right)^{-1}  p\right] \right\} Q( q) \\
	        M( q)
	        &= 
	        Q^\top( q) M_0( q) Q( q).
	    \end{split}
	\end{equation}
\end{prop}

\begin{pf}
	By Proposition \ref{prop:momentumtransformation}, system \eqref{PHS:canonicalMomentumPfaffian} under the momentum transformation \eqref{PHS:momentumTransformationPartition} has the form
	\begin{equation}\label{PHS:nonholreduc:A}
		\begin{split}
			\begin{bmatrix}
				\dot{q} \\
				\dot{\mu} \\
				\dot{p} \\
			\end{bmatrix}
			&=
			\begin{bmatrix}
				0_n & A & Q \\
				-A^\top & \tilde{C}_{11}-\tilde{D}_{11} & -\tilde{C}_{21}^\top-\tilde{D}_{21}^\top \\
				-Q^\top & \tilde{C}_{21}-\tilde{D}_{21} & \tilde{C}_{22}-\tilde{D}_{22} \\
			\end{bmatrix}
			\begin{bmatrix}
				\nabla_q \tilde{\mathcal{H}} \\
				\nabla_\mu \tilde{\mathcal{H}} \\
				\nabla_p \tilde{\mathcal{H}} \\
			\end{bmatrix} \\
			&\qquad+
			\begin{bmatrix}
				0_{n\times m} & 0_{n\times k} \\
				A^\top G_0 & A^\top G_c \\
				Q^\top G_0 & 0_{m\times k} \\
			\end{bmatrix}
			\begin{bmatrix}
				u \\
				\lambda
			\end{bmatrix} \\
			y
			&=
			\begin{bmatrix}
				G_0^\top A & G_0^\top Q \\
				G_c^\top A & G_c^\top Q
			\end{bmatrix}
			\begin{bmatrix}
				\nabla_\mu \tilde{\mathcal{H}} \\
				\nabla_p \tilde{\mathcal{H}}
			\end{bmatrix}.
		\end{split}
	\end{equation}
	Considering the constraint equation \eqref{PHS:PfaffianFactor},
	\begin{equation}\label{PHS:nonholreduc:B}
		\begin{split}
			G_c^\top\dot{ q}
			=
			G_c^\top\nabla_{p_0}\mathcal{H}_0 
			&=
			G_c^\top Q_0^\top\nabla_{\tilde p}\tilde{\mathcal{H}} = 0_{k\times 1}.
		\end{split}
	\end{equation}
	As $G_c^\top A$ is invertible and $G_c^\top Q = 0$, then \eqref{PHS:nonholreduc:B} implies that
	\begin{equation}\label{PHS:nonholreduc:C}
	    \nabla_\mu \tilde{\mathcal{H}} = 0_{k\times 1}.
	\end{equation}
	Considering the $\nabla_{\tilde{p}} \tilde{\mathcal{H}} = \tilde{M}^{-1}\tilde{p}$, we obtain
	\begin{equation}\label{PHS:nonholreduc:D}
		\begin{split}
			\nabla_{\tilde{p}} \tilde{\mathcal{H}}
			&= 
			\left( Q_0^\top M_0 Q_0 \right)^{-1} \tilde{ p} \\
			\begin{bmatrix}
				\mu \\
				 p
			\end{bmatrix}
			&=
			\begin{bmatrix}
				A^\top M_0 A & A^\top M_0 Q \\
				Q^\top M_0 A & Q^\top M_0 Q
			\end{bmatrix}
			\begin{bmatrix}
				\nabla_\mu \tilde{\mathcal{H}} \\
				\nabla_p \tilde{\mathcal{H}}
			\end{bmatrix}.
		\end{split}
	\end{equation}  
	From \eqref{PHS:nonholreduc:C} and \eqref{PHS:nonholreduc:D}, it follows that
	\begin{equation}\label{PHS:nonholreduc:F}
		\begin{split}
		    \nabla_p \tilde{\mathcal{H}}
		    &= 
		    \left( Q^\top M_0 Q \right)^{-1}  p \\
		    \mu 
		    &= 
		    A^\top M_0 Q \left( Q^\top M_0 Q \right)^{-1}  p.
	    \end{split}
	\end{equation}
	Thus $\mu$ is fully determined by $ p$, rendering the $\mu$ dynamics redundant. The modified momentum $\tilde{ p}$ may be computed from the reduced momentum $ p$ using \eqref{PHS:nonholreduc:F} as follows:
	\begin{equation}\label{PHS:nonholreduc:E}
		\begin{split}
			\tilde{ p}
			&= Q_0^\top M_0 Q_0 \nabla_{\tilde{p}} \tilde{\mathcal{H}} \\
			&= Q_0^\top M_0
			\begin{bmatrix}
				A & Q
			\end{bmatrix}
			\begin{bmatrix}
				0_{k\times 1} \\
				\left( Q^\top M_0 Q \right)^{-1}  p
			\end{bmatrix} \\
			&= Q_0^\top M_0 Q \left( Q^\top M_0 Q \right)^{-1}  p.
		\end{split}
	\end{equation}
	Using \eqref{PHS:nonholreduc:E}, consider the Hamiltonian function in \eqref{PHS:TransformedMomentumPfaffian}
	\begin{equation}
		\begin{split}
			\tilde{\mathcal{T}}
			&=
			\frac12 \tilde{ p}^\top \tilde{M}^{-1}( q) \tilde{ p} \\
			&=
			\frac12  p^\top(Q^\top M_0Q)^{-1} p,
		\end{split}
	\end{equation}
	which confirms our choice of Hamiltonian in \eqref{PHS:ReducMomentum} and mass matrix in \eqref{PHS:ReducMomentumMatricies}.
	Combining \eqref{PHS:momentumTransformationPartition} and \eqref{PHS:nonholreduc:E}, the canonical momentum $ p_0$ is given by
	\begin{equation}\label{eq:nonhol:p0_p}
	     p_0
	    = Q_0^{-\top} \tilde{ p}
	    = M_0 Q \left( Q^\top M_0 Q \right)^{-1}  p
	    \text{.}
	\end{equation}	\qed
\end{pf}

The transformation used in \cite{VanDerSchaft1994} can be seen to be a special case of the transformation \eqref{PHS:momentumTransformationPartition} where $A = G_c$. This satisfies the necessary condition that $G_c^\top A = G_c^\top G_c$ be invertible.

An alternative to this transformation arises by choosing $A = M_0^{-1} G_c$.
This satisfies the necessary condition that $G_c^\top A = G_c^\top M_0^{-1} G_c$ is invertible.
From \eqref{PHS:nonholreduc:F}, this choice leads to:
\begin{equation}
    \mu = G_c^\top B \left( B^\top M_0 B \right)^{-1}  p = 0_{k\times 1},
\end{equation}
and the modified mass matrix becomes
\begin{equation}
    \tilde{M} =
    \begin{bmatrix}
        G_c^\top M_0^{-1} G_c & 0_{k\times m} \\
        0_{m\times k} & Q^\top M_0 Q
    \end{bmatrix}.
\end{equation}
This transformation has the property that $\mu$ is equal to the velocities in the directions that the forces due to nonholonomic constraints act, and is thus trivially zero.
As a result of this, the mass matrix is block diagonalised, which further reinforces the point that the decoupled dynamics due to the constraints may be omitted from the model.

Regardless of the choice of a suitable matrix $A$, for isolating and eliminating redundant components of momentum in the presence of nonholonomic constraints, there is still freedom available in the elements of $G_c^\perp$.
For example, we may construct $G_c^\perp$ to render the modified mass matrix constant, as done in \cite{Donaire2015}.

\begin{rem}
	In the derivation of \cite{VanDerSchaft1994}, the term $C$ has the form $-p_0^\top\left[Q_i,Q_j\right]$ where $\left[\cdot,\cdot\right]$ is the lie bracket and $Q_k$ denotes the $k^{th}$ column of $Q$. This expression is equivalent to the form given in \eqref{PHS:ReducMomentumMatricies}.
\end{rem}

\section{PH systems with chained structure}
In this section, two configuration transformations are proposed for the system \eqref{PHS:ReducMomentum}. The first transformation $f_z:q\to z$ is used to transform the system such that the transformed system has a \textit{chained structure}, a generalisation of chained form. A second discontinuous configuration transformation $f_w:z\to w$ is proposed which serves two purposes: Firstly, the asymptotic behaviour of $z(t)$ is reduced to the asymptotic behaviour of a single variable in the $w$ space. Secondly, the control objective can be addressed by shaping the potential energy to be quadratic in $w$.

\subsection{Chained structure}
Chained form systems are two input kinematic systems described by the equations:
\begin{equation}\label{PHS:chainedForm}
	\begin{bmatrix}
		\dot{z}_1 \\ \dot{z}_2 \\ \dot{z}_3 \\ \dot{z}_4 \\ \vdots \\ \dot{z}_n
	\end{bmatrix}
	=
	\underbrace{
		\begin{bmatrix}
			1 & 0 \\
			0 & 1 \\
			z_2 & 0 \\
			z_3 & 0 \\
			\vdots & \vdots \\
			z_{n-1} & 0
		\end{bmatrix}
		}_{Q_c(z)}
	\begin{bmatrix}
		v_1 \\ v_2
	\end{bmatrix},
\end{equation}
where $v_1, v_2 \in \mathbb{R}$ are velocity inputs and $z_i\in\mathbb{R}$ are configuration variables \cite{Bloch2003,Astolfi1996,Murray1991a}. The kinematic models of many interesting nonholonomic systems can be expressed in chained form under the appropriate coordinate and input transformations. A procedure to transform kinematic models into chained form was presented in \cite{Murray1991a}.

Consider now a new set of generalised coordinates $z = f_z(q)$ for the system \eqref{PHS:ReducMomentum} where $f_z$ is invertible. 
By Lemma 2 of \cite{Fujimoto2001a}, the system \eqref{PHS:ReducMomentum} is equivalently described in the coordinates $(z,p)$ by:
	\begin{equation}\label{PHS:ReducMomentumZCoord}
		\begin{split}
			\begin{bmatrix}
				\dot{z} \\\dot{ p} \\ 
			\end{bmatrix}
			&=
			\begin{bmatrix}
				0_n & Q_z \\
				-Q_z^\top & C_z-D_z
			\end{bmatrix}
			\begin{bmatrix}
				\nabla_z \mathcal{H}_z \\ 
				\nabla_p \mathcal{H}_z
			\end{bmatrix}
			+
			\begin{bmatrix}
				0_{n\times m} \\ 
				G_z \\ 
			\end{bmatrix}
			u \\
			y
			&=
			G_z^\top\nabla_p \mathcal{H}_z \\
			\mathcal{H}_z 
			&= 
			\frac{1}{2} p^\top M_z^{-1} p + \mathcal{V}_z(z),
		\end{split}
	\end{equation}
	where 
	\begin{multicols}{2}
		\noindent
		\begin{equation*}
			\begin{split}
				\mathcal{H}_z &= \mathcal{H}( q, p)|_{ q = f_z^{-1}(z)} \\
				\mathcal{V}_z &= \mathcal{V}( q)|_{ q = f_z^{-1}(z)} \\
				M_z &= M( q)|_{ q = f_z^{-1}(z)} \\
				Q_z &= \frac{\partial}{\partial q}(f_z) Q(q)|_{ q = f_z^{-1}(z)} \\
			\end{split}
		\end{equation*}
		\begin{equation}\label{CbI:ControllerMatricies}
			\begin{split}
				C_z &= C( q, p)|_{ q = f_z^{-1}(z)} \\
				D_z &= D( q, p)|_{ q = f_z^{-1}(z)} \\
				G_z &= G( q)|_{ q = f_z^{-1}(z)}.
			\end{split}
		\end{equation}
	\end{multicols}
	\noindent
Considering the nonholonomic system expressed in the coordinates $(z,p)$ given by \eqref{PHS:ReducMomentumZCoord}, pH systems with chained structure can now be defined.

\begin{defn}\label{PHS:chainedDefn}
	A nonholonomic pH system of the form \eqref{PHS:ReducMomentumZCoord} has a \textit{chained structure} if $Q_z$ has a left annihilator of the form 
	\begin{equation}\label{PHS:chainedNullspace}
		Q_z^\perp(z)
		=
		\begin{bmatrix}
			-z_2 & 0 & 1 & 0 & \cdots & 0 \\
			-z_3 & 0 & 0 & 1 & \cdots & 0 \\
			\vdots & \vdots & \vdots & \vdots & \ddots & \vdots \\
			-z_{n-1} & 0 & 0 & 0 & \cdots & 1
		\end{bmatrix}
		\in\mathbb{R}^{(n-2)\times n}.
	\end{equation}
\end{defn}

The relationship between chained systems and chained structure is now apparent; $Q_c$ as defined in \eqref{PHS:chainedForm} has the trivial left annihilator of the form \eqref{PHS:chainedNullspace}. This annihilator is then used as the defining property used in our definition of chained structure. By this definition, pH systems with chained structure are two-input systems ($u\in\mathbb{R}^2$) with momentum space of dimension 2 ($p_r\in\mathbb{R}^2$).

\begin{rem}
	The kinematics associated with \eqref{PHS:ReducMomentum} are $\dot q = Q\nabla_{p_r}\mathcal{H}_r$ where $\nabla_{p_r}\mathcal{H}_r$ is considered an input to the kinematic system. If this kinematic system admits a feedback transformation that transforms it into chained form using the method presented in \cite{Murray1991a}, then by Proposition 1 of \cite{Fujimoto2012}, there exists a coordinate and momentum transformation that transforms \eqref{PHS:ReducMomentum} into \eqref{PHS:ReducMomentumZCoord} with $Q_z = Q_c$. Such a system clearly has a chained structure.
\end{rem}

\subsection{Discontinuous coordinate transformation}
A discontinuous coordinate transformation $f_w:z\to w$ for systems with chained structure is now proposed. The purpose of this transformation is to render the open-loop dynamics in a form whereby the control problem can be addressed by shaping the potential energy to be quadratic in $w$.

The transformation $f_w$ is defined implicitly by its inverse mapping:
\begin{equation}\label{PES:CoordinateTransform}
	\begin{split}
		\begin{bmatrix}
			z_1 \\ z_2 \\ z_3 \\ \vdots \\ z_j \\ \vdots \\ z_n
		\end{bmatrix}
		=
		f_w^{-1}(w)
		=
		\begin{bmatrix}
			w_1 \\
			w_1w_2 + \sum_{i=3}^{n}\frac{1}{(i-2)!}w_1^{(i-2)}w_{i} \\
			\sum_{i=3}^{n}\frac{1}{(i-1)!}w_1^{(i-1)}w_{i} \\
			\vdots \\
			\sum_{i=3}^{n}\frac{1}{(i+j-4)!}w_1^{(i+j-4)}w_{i} \\
			\vdots \\
			\sum_{i=3}^{n}\frac{1}{(i+n-4)!}w_1^{(i+n-4)}w_{i}
		\end{bmatrix}.
	\end{split}
\end{equation}
The inverse transformation \eqref{PES:CoordinateTransform} has been constructed to satisfy two properties. Firstly, it can be seen that the mapping $f_w^{-1}$ is smooth and if $w_1 = 0$ then $z = 0$. Thus the control problem can be addressed in the $w$ coordinates simply by controlling $w_1$. The second useful property of \eqref{PES:CoordinateTransform} is that each element of $z(w) = f_w^{-1}(w)$ satisfies the relationship
\begin{equation}
	z_{i+1}(w) = \int z_i(w)|_{w_2 = 0}dw_1
\end{equation}
for $i \geq 2$. Considering chained form systems \eqref{PHS:chainedForm}, such a definition is closely related to the underlying system but integration now occurs spatially, rather than temporally.

The remainder of this section is devoted to proving that $f_w:z\to w$, defined implicitly by \eqref{PES:CoordinateTransform}, is well defined for all $z_1\neq 0$.
\begin{lem}\label{InvertSn}
	The matrix $S_n$ defined as
	\begin{equation}\label{Sn}
		\begin{split}
			S_n
			=
			\begin{bmatrix}
				\frac{1}{2!} & \frac{1}{3!} & \cdots & \frac{1}{(n-1)!} \\
				\frac{1}{3!} & \frac{1}{4!} & \cdots & \frac{1}{n!} \\
				\vdots & \vdots & \ddots & \vdots \\
				\frac{1}{(n-1)!} & \frac{1}{n!} & \cdots & \frac{1}{(2n-4)!} \\
			\end{bmatrix}
		\end{split}
	\end{equation}
	is invertible for all $n \geq 3$.
\end{lem}

\begin{pf}
	The proof is provided in the Appendix\footnote{The proof of Lemma \ref{InvertSn} was proposed by \textit{user1551} on \textit{math.stackexchange.com}}.\qed
\end{pf}
\begin{prop}
	The function $f_w:z\to w$, defined implicitly by \eqref{PES:CoordinateTransform}, is well defined for all $z_1\neq 0$.
\end{prop}

\begin{pf}
	First note that $z_1 = w_1$ is invertible for all $z_1$. Let
	$z' = \begin{bmatrix} z_3 & \cdots & z_n \end{bmatrix}^\top$, $w' = \begin{bmatrix} w_3 & \cdots & w_n \end{bmatrix}^\top$ and $N(w_1) = \operatorname{diag}(w_1,w_1^2,\dots,w_1^{n-2})$.
	$z'$ and $w'$ are related by
	\begin{equation}\label{control:ZpTransformation}
		z' = N(w_1)S_nN(w_1)w'.
	\end{equation}
	$N(w_1)$ is invertible for all $w_1\neq 0$ and $S_n$ is invertible by Lemma \ref{InvertSn}. Thus, we have that
	\begin{equation}
		w' = N^{-1}(z_1)S_n^{-1}N^{-1}(z_1)z'.
	\end{equation}
	Finally, the transformation for $w_2$ can be solved algebraically as the solution to
	\begin{equation}\label{w2}
		w_2 = \frac{z_2}{z_1} - \sum_{i=3}^{n}\frac{1}{(i-3)!}z_1^{(i-2)}w_{i}(z).
	\end{equation}\qed
\end{pf}

By Lemma 2 of \cite{Fujimoto2001a}, \eqref{PHS:ReducMomentumZCoord} can be equivalently described in the coordinates $(w,p)$ by:
\begin{equation}\label{wDynamics}
	\begin{split}
		\begin{bmatrix}
			\dot{w} \\ \dot{p} \\ 
		\end{bmatrix}
		&=
		\begin{bmatrix}
			0_n & Q_w \\
			-Q_w^\top & C_w-D_w
		\end{bmatrix}
		\begin{bmatrix}
			\nabla_w \mathcal{H}_w \\
			\nabla_p \mathcal{H}_w \\ 
		\end{bmatrix}
		+
		\begin{bmatrix}
			0_{n\times 2} \\
			G_w \\ 
		\end{bmatrix}
		u \\
		\mathcal{H}_w
		&= 
		\frac{1}{2}p^\top M_w^{-1}p + \mathcal{V}_w,
	\end{split}
\end{equation}
where 
\begin{multicols}{2}
	\noindent
	\begin{equation*}
		\begin{split}
			\mathcal{H}_w &= \mathcal{H}_z( z, p)|_{ z = f_w^{-1}(w)} \\
			\mathcal{V}_w &= \mathcal{V}_z( z)|_{ z = f_w^{-1}(w)} \\
			M_w &= M_z( z)|_{ z = f_w^{-1}(w)} \\
			Q_w &= \frac{\partial}{\partial z}(f_w) Q_z(z)|_{ z = f_w^{-1}(w)} \\
		\end{split}
	\end{equation*}
	\begin{equation}\label{QwDefs}
		\begin{split}
			C_w &= C_z( z, p)|_{ z = f_w^{-1}(w)} \\
			D_w &= D_z( z, p)|_{ z = f_w^{-1}(w)} \\
			G_w &= G_z( z)|_{ z = f_w^{-1}(w)}.
		\end{split}
	\end{equation}
\end{multicols}

\section{Stabilisation via potential energy shaping and damping injection}\label{PES}
In this section, a discontinuous control law is proposed for the nonholonomic pH system \eqref{PHS:canonicalMomentumPfaffian}. \\

\begin{assum}\label{AssumpChained}
	Consider the nonholonomic pH system \eqref{PHS:canonicalMomentumPfaffian} that has been expressed without Lagrange multipliers in the form \eqref{PHS:ReducMomentum}.
	It is assumed that there exists a coordinate transformation $f_z:q\to z$ such that $f_z^{-1}(0_{n\times 1}) = 0_{n\times 1}$ and the dynamics expressed as a function of $z$, given by \eqref{PHS:ReducMomentumZCoord}, have a chained structure with smooth $Q_z$. \\
\end{assum}

Under Assumption \ref{AssumpChained}, \eqref{PHS:canonicalMomentumPfaffian} can be equivalently represented in the $(z,p)$ coordinates by \eqref{PHS:ReducMomentumZCoord} with a chained structure or in the coordinates $(w,p)$ as per \eqref{wDynamics}. 

\subsection{Stabilising control law}

\begin{prop}
	Consider the system \eqref{wDynamics} in closed-loop with the control law
	\begin{equation}\label{CtrlLaw}
		\begin{split}
			u &=
			-G_w^{-1}
			\left\lbrace
			Q_w^\top\left[Lw - \nabla_w\mathcal{V}_w\right] \right.\\
			&\left.\phantom{------}
			+	
			\left[\hat D + \frac{k}{w_1^2}Q_w^\top e_1e_1^\top Q_w\right]\nabla_p\mathcal{H}_w
			\right\rbrace,
		\end{split}
	\end{equation}
	where $\hat D\in \mathbb{R}^{2\times 2}$ is positive definite, $e_1\in\mathbb{R}^n$ is the first standard basis vector, $k>0$ is a constant and $L = \operatorname{diag}(l_1,\dotsc,l_n)$ is a constant positive matrix.
	The closed-loop dynamics have the form
	\begin{equation}\label{CLDynamics}
		\begin{split}
			\begin{bmatrix}
				\dot{w} \\ \dot{p}
			\end{bmatrix}
			&=
			\begin{bmatrix}
				0_n & Q_w \\
				-Q_w^\top & C_w-D_d
			\end{bmatrix}
			\begin{bmatrix}
				\nabla_w \mathcal{H}_d \\ 
				\nabla_p \mathcal{H}_d
			\end{bmatrix} \\
			\mathcal{H}_d
			&= 
			\frac{1}{2}p^\top M_w^{-1}p + \underbrace{\frac12 w^\top Lw}_{\mathcal{V}_{d}},
		\end{split}
	\end{equation}
	where $D_d = D_w + \hat D + \frac{k}{w_1^2}Q_w^\top e_1e_1^\top Q_w$.
\end{prop}

\begin{pf}
	The proof follows from direct computation. \qed
\end{pf}

The proposed control law is comprised of two parts: potential energy shaping and damping injection. The term $-G_w^{-1}Q_w^\top\left[Lw - \nabla_w\mathcal{V}_w\right]$ can be considered to be potential energy shaping as its role is to replace the potential energy term $\mathcal{V}_w$ of \eqref{wDynamics} with $\mathcal{V}_{d}$. The role of the potential energy shaping is to drive the system to the configuration $z_1 = w_1 = 0$ whilst keeping each $w_i$ bounded. Likewise, the term $-G_w^{-1}\left[\hat D + \frac{k}{w_1^2}Q_w^\top e_1e_1^\top Q_w\right]\nabla_p\mathcal{H}_w$ can be considered damping injection as it increases the damping from $D_w$ to $D_d$. As the dynamics \eqref{wDynamics} are not defined at $z_1 = w_1 = 0$, the role of the damping injection is to ensure that the system cannot reach the configuration $z_1 = 0$ in finite time. The combination of the two terms drives the system to the configuration $w_1 = 0$ asymptotically, but prevents any finite time convergence.

\begin{figure}[htbp]
	\centering
	\includegraphics[trim = 0mm 0mm 0mm 0mm, clip, width=1.0\linewidth]{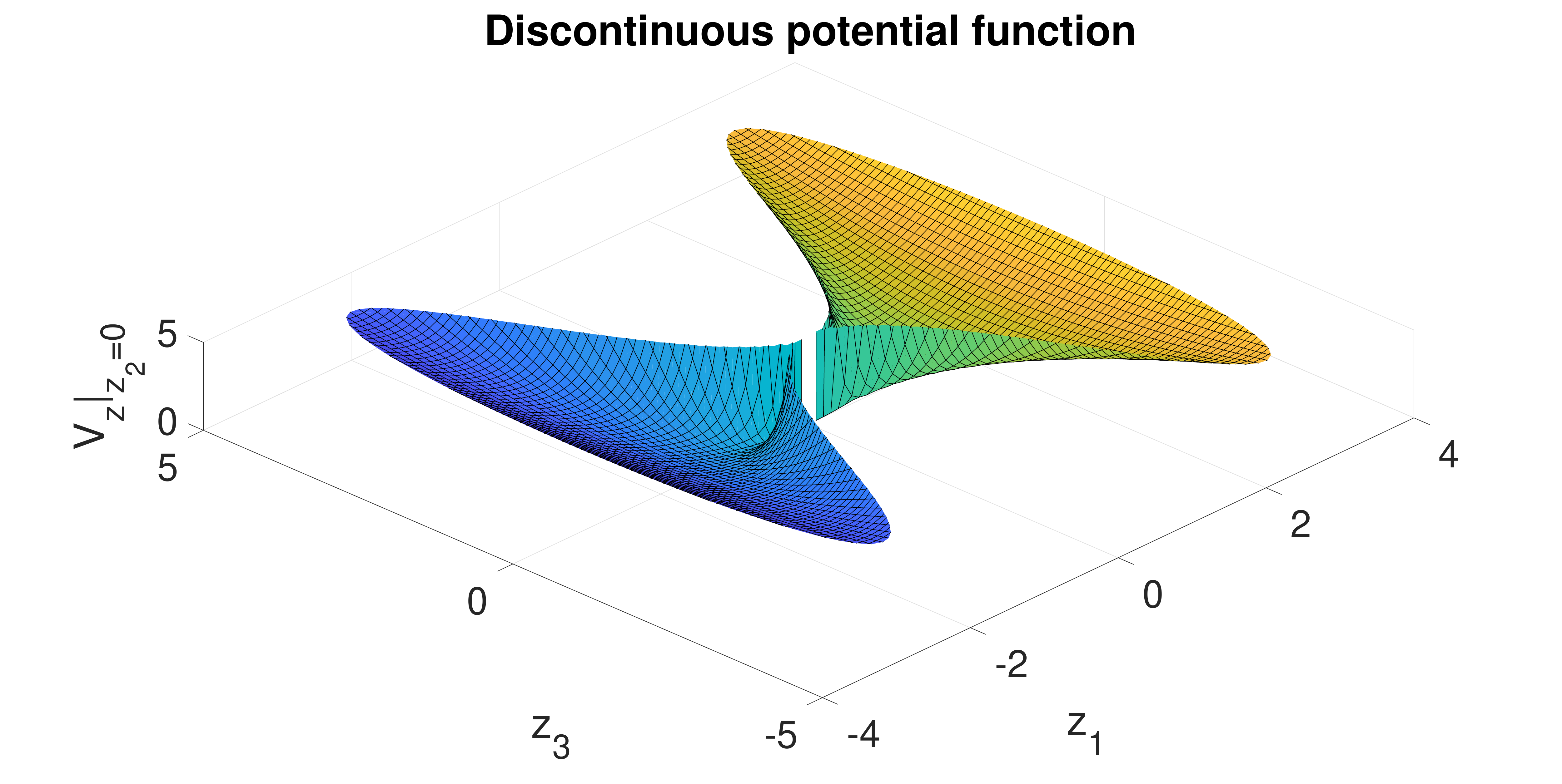}
	\caption{A component of the discontinuous potential energy function $\mathcal{V}_w = \frac12 w^\top L w$ with $n=3$ and $L = I_{3\times 3}$ expressed as a function of $z$ projected onto $z_2 = 0$.}
	\label{PotFunc}
\end{figure}

To visualise the potential function $\mathcal{V}_{dw}$, consider the case that $n=3$. The resulting discontinuous transformation is given by
\begin{equation}
	\begin{split}
		f_w(z)
		=
		\begin{bmatrix}
			z_1 \\ \frac{z_2}{z_1}-\frac{2z_3}{z_1^2} \\ \frac{2z_3}{z_1^2}
		\end{bmatrix}.
	\end{split}
\end{equation}
Figure \ref{PotFunc} is a plot of the function
\begin{equation}
	\begin{split}
		V_z(z)
		&=
		\frac12 w_1(z)^2 + \frac12 w_2(z)^2|_{z_2=0} + \frac12 w_3(z)^2 \\
		&=
		\frac12 z_1^2 + \frac{4z_3^2}{z_1^4},
	\end{split}
\end{equation}
which is part of the shaped potential energy function, projected onto $z_2 = 0$. Interestingly, the level sets resemble ``figure of eights" and the function diverges as $z_1$ tends to $0$, unless the ratio $\frac{z_3}{z_1^2}$ is bounded. Thus, it can be seen that the potential function only allows the system to approach the origin from particular directions.

\begin{rem}
	As $Q_w$ is full rank, the term $\nabla_p\mathcal{H}_w$ can be expressed as a function of $w,\dot w$. Thus, the control law \eqref{CtrlLaw} can be expressed independent of the systems mass matrix $M_w$. Further, the control law is independent of the open-loop damping structure $D_w$. Thus the proposed control scheme is robust against damping parameters. This is similar to the case of energy shaping of fully-actuated mechanical systems.
\end{rem}
\begin{rem}
The control law has been presented as a function of $(w,p)$ and stability analysis will be performed primarily in these coordinates. However, the control law \eqref{CtrlLaw} can be equivalently expressed as a function of $(z,p)$ via the mapping $f_w^{-1}$:
\begin{equation}\label{CtrlLawZ}
	\begin{split}
		u &=
		-G_z^{-1}
		\left\lbrace
		Q_z^\top\left[\frac{\partial^\top}{\partial z}(f_w) \nabla_z \mathcal{V}_{dz} - \nabla_z\mathcal{V}_z\right] \right.\\
		&\left.\phantom{------}
		+	
		\left[\hat D + \frac{k}{z_1^2}Q_z^\top e_1e_1^\top Q_z\right]\nabla_p\mathcal{H}_z
		\right\rbrace,
	\end{split}
\end{equation}
where $\mathcal{V}_{dz} = \mathcal{V}_{dw}|_{w = f_w(z)}$. The control law is discontinuous as a function of $z$ due to the terms $\frac{\partial^\top}{\partial z}(f_w) \nabla_z \mathcal{V}_{dz}$ and $\frac{k}{z_1^2}Q_z^\top e_1\dot{z}_1$.
\end{rem}

\subsection{Stability analysis}
The remainder of this section is devoted to showing that the closed-loop dynamics \eqref{CLDynamics} are well defined for all time and $w_1\to 0$ as $t\to \infty$. To do this, let $x = (p,w)$ and define the set 
\begin{equation}\label{SetU}
	U = \{x|\mathcal{H}_d(x)\leq\mathcal{H}_d(0),w_1 \neq 0\}.
\end{equation}
Recalling that $f_w$ is well defined for $z_1 \neq 0$, the closed-loop dynamics \eqref{CLDynamics} are well defined on $U$.

The following proposition demonstrates that the set $U$ is positively invariant which implies that provided that the system is initialised with $w_1(t_0)\neq 0$, then \eqref{CLDynamics} describes the system dynamics for all time.
\begin{lem}\label{FuncInequality}
	Any real valued function $f(x)$ satisfies the inequality,
	\begin{equation}
		-\frac{1}{x_2-x_1}\left(\int_{x_1}^{x_2}f(x)dx\right)^2
		\geq
		-\int_{x_1}^{x_2}f^2(x)dx,
	\end{equation}
	where $x_2 > x_1$ are in the domain of $f$.
\end{lem}
\begin{pf}
	The proof follows from the Schwarz inequality \cite{lieb2001analysis}. Details are provided in the Appendix. \qed
\end{pf}

\begin{prop}\label{z1neq0}
	If the closed-loop dynamics \eqref{CLDynamics} have initial conditions such that $w_1(t_0) \neq 0$, then the set $U$ is positively invariant. That is, $x(t)\in U$ for all $t \geq t_0$.
\end{prop}
\begin{pf}
	The time derivative of $\mathcal{H}_{d}$ satisfies
	\begin{equation}\label{HamDeriv}
		\dot{\mathcal{H}}_{d}
		=
		-p^\top M_w^{-1}D_dM_w^{-1}p \leq 0.
	\end{equation}
	For any time interval $\Delta t = [t_0,T)$ with the property that $w_1(t)\neq 0\ \forall t\in\Delta t$, the shaped Hamiltonian will satisfy $\mathcal{H}_d(t)\leq\mathcal{H}_d(t_0)$. Considering that $\mathcal{H}_d$ is quadratic in $p$ and $w$, this means that $p(t)$ and $w(t)$ are bounded for all $t\in\Delta t$. This means that $z(t)$ is bounded on $\Delta t$ as $f_w^{-1}$ is smooth \eqref{PES:CoordinateTransform}. As $w_1 = z_1$, $p(t)$ is bounded and $Q_z$ is smooth, considering the dynamics \eqref{PHS:ReducMomentumZCoord} reveal that $\dot w_1(t) = \dot z_1(t)$ is bounded for all $t\in\Delta t$. As $\dot w(t)$ is bounded, $\lim_{t\to T}w_1(t)$ exists for any $T$.

	Now, for the sake of contradiction, assume that $\lim_{t\to T}w_1(t) = 0$ for some finite $T\in [t_0,\infty)$. Taking any interval $[t_1,T]$, such that $t_1 \geq t_0$, pick $t'$ such that it maximises $w_1^2(t)$ on the interval $[t_1,T]$. The time derivative of the Hamiltonian satisfies
	\begin{equation}
		\begin{split}
			\dot{\mathcal{H}}_{d}(t) 
			&\leq 
			-\frac{k}{w_1^2}\nabla_{p}^\top\mathcal{H}_{d}Q_w^\top e_1e_1^\top Q_w\nabla_{p}\mathcal{H}_{d} \\
			&=
			-\frac{k}{w_1^2(t)}\dot{w}_1^2(t).
		\end{split}
	\end{equation}
	Integrating with respect to time from $t'$ to $T$
	\begin{equation}
		\begin{split}
			\mathcal{H}_{d}(T) - \mathcal{H}_{d}(t')
			&\leq 
			-\int_{t'}^{T}\frac{k}{w_1^2(t)}\dot{w}_1^2(t) dt.
		\end{split}
	\end{equation}
	As $w_1(t') = \max\{w_1(t)\}\ \forall t\in[t_1,T]$,
	\begin{equation}
		\begin{split}
			\mathcal{H}_{d}(T) - \mathcal{H}_{d}(t')
			&\leq 
			-\frac{k}{w_1^2(t')}\int_{t'}^{T}\dot{w}_1^2(t) dt.
		\end{split}
	\end{equation}
	Applying Lemma \ref{FuncInequality} to this inequality
	\begin{equation}
		\begin{split}
			\mathcal{H}_{d}(T) &- \mathcal{H}_{d}(t') \\
			&\leq 
			-\frac{k}{w_1^2(t')}\frac{1}{T-t'}\left(\int_{t'}^{T}\dot{w}_1(t) dt\right)^2 \\
			&\leq 
			-\frac{k}{w_1^2(t')}\frac{1}{T-t'}\left(w_1(T) - w_1(t')\right)^2 \\
			&\leq 
			-\frac{k}{w_1^2(t')}\frac{1}{T-t'}w_1^2(t') \\
			&\leq 
			-\frac{k}{T-t'}.
		\end{split}
	\end{equation}
	As $T - t' \leq T - t_1$ is arbitrarily small, the right hand side of this inequality can be made arbitrarily large by choosing $T-t_1$ small enough. However, the Hamiltonian is lower bounded, thus we have a contradiction. Thus, we conclude that there is no finite $T$ such that $\lim_{t\to T}z_1(t) = 0$ which implies that $U$ is positively invariant. \qed
\end{pf}

By Proposition \ref{z1neq0}, it is clear that the closed-loop dynamics are well defined for all finite time $t<\infty$. The asymptotic behaviour of the system is now considered. The underlying approach taken here is to show that the system cannot approach any subset of $U$ asymptotically. To this end, the following Lemma shows that $\dot{\mathcal{H}}_d$ cannot be identically equal to zero on the set $U$.

\begin{lem}\label{NoInvarSet}
	Consider the closed-loop dynamics \eqref{CLDynamics} defined for all $w_1\neq 0$. On the set $U$ there is no solution to \eqref{CLDynamics} satisfying $\dot{\mathcal{H}}_d = 0$ identically.
\end{lem}

\begin{pf}
	The time derivative of $\mathcal{H}_{d}$ along the trajectories of \eqref{CLDynamics} are given by \eqref{HamDeriv}.
	As $D_d, M_w > 0$, for \eqref{HamDeriv} to be identically equal to zero, $p$ must be identically equal to zero. This means that $\dot{p} = 0_{2\times 1}$ along such a solution.
	
	Evaluating the $\dot{p}$ dynamics of \eqref{CLDynamics} at $p = \dot{p} = 0_{2\times 1}$ results in
	\begin{equation}\label{InvarSetDetect}
		-Q_w^\top Lw
		= 
		0_{n\times 1}.
	\end{equation}
	From \eqref{QwDefs}, $Q_w^\top = \left[Q_z^\top(z)\frac{\partial^\top}{\partial z}(f_w)\right]\big|_{z = f_w^{-1}(w)}$, which allows \eqref{InvarSetDetect} to be rewritten as
	\begin{equation}\label{detect2}
		Q_z^\top(z)\big|_{z = f_w^{-1}(w)}\underbrace{\frac{\partial^\top}{\partial z}(f_w)\big|_{z = f_w^{-1}(w)} Lw}_{Q_{ra}}
		= 
		0_{n\times 1}.
	\end{equation}
	The expression \eqref{detect2} is satisfied if the columns of $Q_{ra}$ are in the null-space of $Q_z^\top$. Letting $Q_z^\perp$ be any full rank left annihilator of $Q_z$, \eqref{detect2} is satisfied if
	\begin{equation}\label{control:smoothConvergence:2}
		\frac{\partial^\top}{\partial z}(f_w)|_{z = f_w^{-1}(w)} Lw = \left[Q_z^\perp\big|_{z = f_w^{-1}(w)}\right]^\top a(w),
	\end{equation}
	where $a(w)\in\mathbb{R}^{n-2}$ is an arbitrary vector.
	Rearranging \eqref{control:smoothConvergence:2} results in
	\begin{equation}\label{control:smoothConvergence:3}
		\begin{split}
			Lw &= \frac{\partial^\top}{\partial w}(f_w^{-1}) \left[Q_z^\perp\big|_{z = f_w^{-1}(w)}\right]^\top a(w).
		\end{split}
	\end{equation}
	Taking $Q_z^\perp$ to be \eqref{PHS:chainedNullspace}, the term $\frac{\partial^\top}{\partial w}(f_w^{-1})\left[Q_z^\perp\right]^\top$ is expanded in \eqref{control:smoothConvergence:4}, where $\ast$ denotes an unevaluated element. 
	\begin{figure*}[!t]
		\small
		\begin{equation}\label{control:smoothConvergence:4}
			\underbrace{
			\begin{bmatrix}
				1 & w_2+\sum_{i=3}^{n}\frac{1}{(i-3)!}w_1^{(i-3)}w_{i} & z_2(w)|_{w_2=0} & \cdots & z_{n-1}(w) \\
				0 & w_1 & 0 & \cdots & 0 \\
				0 & w_1 & \frac{1}{2!} w_1^2 & \cdots & \frac{1}{(n-1)!}w_1^{n-1} \\
				0 & \frac{1}{2!}w_1^{2} & \frac{1}{3!} w_1^3 & \cdots & \frac{1}{n!}w_1^{n} \\
				\vdots & \vdots & \vdots & \ddots & \vdots \\
				0 & \frac{1}{(n-2)!}w_1^{n-2} & \frac{1}{(n-1)!}w_1^{n-1} & \cdots & \frac{1}{(2n-4)!}w_1^{(2n-4)}
			\end{bmatrix}}_{\frac{\partial^\top}{\partial w}(f_w^{-1})}
			\underbrace{
			\begin{bmatrix}
				-z_2 & -z_3 & \cdots & -z_{n-1} \\
				0 & 0 & \cdots & 0 \\
				1 & 0 & \cdots & 0 \\
				0 & 1 & \cdots & 0 \\
				\vdots & \vdots & \ddots & \vdots \\
				0 & 0 & \cdots & 1
			\end{bmatrix}}_{\left[Q_z^\perp\right]^\top}
			=
			\begin{bmatrix}
				-w_1w_2 & 0 & \cdots & 0 \\
				0 & 0 & \cdots & 0 \\
				\ast & \ast & \cdots & \ast \\
				\ast & \ast & \cdots & \ast \\
				\vdots & \vdots & \ddots & \vdots \\
				\ast & \ast & \cdots & \ast \\
			\end{bmatrix}
		\end{equation}
		\normalsize
		\hrulefill
		\vspace*{4pt}
	\end{figure*}
	
	Considering the second row of \eqref{control:smoothConvergence:3} with the evaluation in \eqref{control:smoothConvergence:4}, $w_2 = 0$. Substituting in $w_2 = 0$ and considering the first row of \eqref{control:smoothConvergence:3}, $w_1 = 0$. Clearly, such a solution is not contained in $U$. \qed
\end{pf}

When analysing the asymptotic behaviour of Hamiltonian systems, it is typical to invoke LaSalle's invariance principle to show that the system converges to the largest invariant set contained within $\dot{\mathcal{H}}_{d} = 0$. However, we note that as the closed-loop dynamics \eqref{CLDynamics} have a discontinuous right hand side, LaSalle's theorem does not apply. 

The following Proposition shows that $w_1$ does indeed tend towards zero. The intuition behind that Proposition is noticing that if the system were to converge to a set that is at least partially contained within $U$, then $\dot{\mathcal{H}}_d$ would be identically equal to zero on this set. Note that the proof presented here is very similar in nature to the proof of LaSalle's theorem found in \cite{Khalil1996}.

\begin{prop}\label{w1Zero}
	If the closed-loop dynamics \eqref{CLDynamics} have initial conditions such that $w_1(t_0) \neq 0$ then $\lim_{t\to\infty} w_1 = 0$. Furthermore, this implies that $\lim_{t\to\infty} z = 0_{n\times 1}$.
\end{prop}
\begin{pf}
	Recall that $x = (p,w)$ and $U$ defined in \eqref{SetU} is positively invariant by Proposition \ref{z1neq0}. As $\mathcal{H}_d$ is quadratic in $p$ and $w$, it is radially unbounded which implies that $U$ is a bounded set.

	By the Bolzano-Weierstrass theorem, any solution $x(t)$ admits an accumulation point as $t\to\infty$. The set of all accumulation points is denoted $L^+$. To see that $x(t) \to L^+$, first presume that does not. Then, there exists a sequence $t_k$ such that $d(x(t_k),L^+) > \epsilon$. As $x(t)$ is bounded, $x(t_k)$ has a convergent subsequence by the Bolzano-Weierstrass theorem and such a subsequence converges to $L^+$, which is a contradiction. 
	
	As $\mathcal{H}_d(t)$ is monotonically decreasing and bounded below by zero, $\lim_{t\to\infty}\mathcal{H}_{d} = \mathcal{H}_L$ exists. Now suppose that $V = L^+\cap U\neq\emptyset$. By definition, for each $y\in V$, there exists a sequence $t_n$ such that $\lim_{n\to \infty}x(t_n) = y$. As $\mathcal{H}_d$ is continuous and $\lim_{t\to\infty}\mathcal{H}_d = \mathcal{H}_L$, $\mathcal{H}(V) = \mathcal{H}_L$. By the continuity of solutions on $U$ and Lemma \ref{z1neq0}, a solution $x(t)$ with $x(0) = y$ is contained in $V$. Thus, such a solution satisfies $\dot{\mathcal{H}}_d(t) = 0$.
	
	But by Proposition \ref{NoInvarSet}, there is no solution in the set $U$ satisfying $\dot{\mathcal{H}}_d = 0$ identically. Thus we conclude that $V=\emptyset$ and $L^+$ is contained in the set 
	\begin{equation}
		\bar{U}\setminus U = \{x|\mathcal{H}(x)\leq\mathcal{H}(0),w_1 = 0\}.
	\end{equation}
	As $x(t)\to L^+$, $w_1(t)\to 0$.
	
	Considering the coordinate transformation \eqref{PES:CoordinateTransform}, and noting that each $w_i(t)$ is bounded, $w_1(t)\to 0$ implies that each $z_i(t)$ tends towards zero. \qed
\end{pf}

Notice that although $z$ tends towards the origin asymptotically, the asymptotic behaviour of $p$ has not been established. Clearly $p(t)\in\mathcal{L}_\infty$ as $\frac{1}{2}p^\top M_w^{-1}p < \mathcal{H}_d(t) \leq \mathcal{H}_d(0)$ for all time. Further analysis is considered beyond the scope of this paper and left an area for future research. 

\section{Car-like system example}\label{car}
In this section, a car-like system is modelled and controlled. The system is shown to have a chained structure and thus is able to be controlled with the control law \eqref{CtrlLaw}.

\subsection{Modelling the car-like system}
\begin{figure}
	\centering
	\includegraphics[trim = 3mm 4mm 3mm 4mm, clip, width=.6\linewidth]{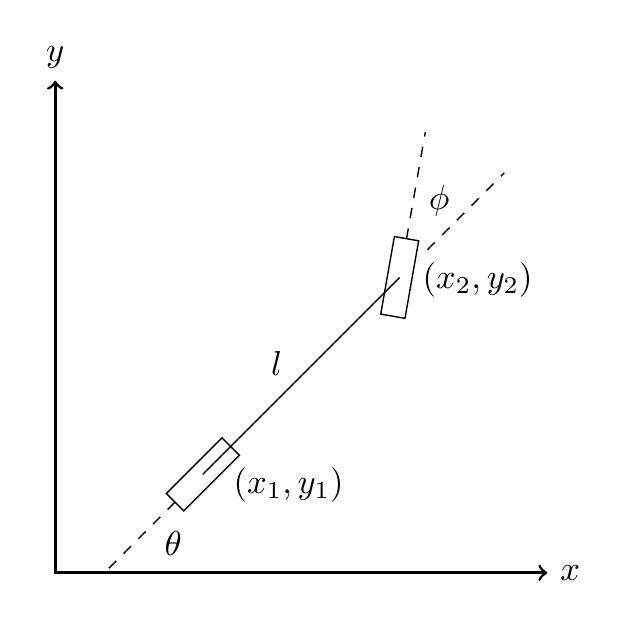}
	\caption{The car-like system is fixed to the ground at the points $(x_1,y_1)$ and $(x_2,y_2)$. The point $(x_1,y_1)$ is able to move along the direction of $\theta$ while $(x_2,y_2)$ can move in the direction of $\theta+\phi$. The two wheels have a fixed distance given by $l$ and the rear wheel is unable to move relative to the frame. The front wheel can pivot about its point of anchor to the frame. We have two control inputs to this system; A torque can be applied to the front wheel to cause it to turn about $\phi$ and a force can be applied to the rear wheel along the direction given by $\theta$.}
	\label{car:image}
\end{figure}
The car-like system (Figure \ref{car:image}) can be modelled as a mechanical pH system \eqref{PHS:canonicalMomentumPfaffian}, subject to two Pfaffian constraints. The kinetic co-energy of the system is computed as
\begin{equation}\label{car:KineticEnergy}
	\begin{split}
		\mathcal{T}^* = \frac12 m_1(\dot{x}_1^2 + \dot{y}_1^2) + \frac12 J_1\dot{\theta}_1^2 + \frac12 m_2(\dot{x}_2^2 + \dot{y}_2^2) + \frac12 J_2\dot{\theta}_2^2,
	\end{split}
\end{equation}
where $m_i, J_i$ are the masses and moment of inertias of the rear and front wheels respectively. The system is subject to two holonomic constraints
\begin{equation}\label{car:holonomicConstraints}
	\begin{split}
		x_2 &= x_1 + l\cos\theta \\
		y_2 &= y_1 + l\sin\theta,
	\end{split}
\end{equation}
which must be satisfied along any solution to the system. The need for these auxiliary equations can be removed by the appropriate selection of configuration variables. 

Our objective is to stabilise the configuration of the rear wheel $(x_1,y_1,\theta)$ to the origin. As such, the coordinates $x_2$ and $y_2$ are eliminated from our dynamic equations by using the identities \eqref{car:holonomicConstraints}. Taking the time derivatives of the constraints \eqref{car:holonomicConstraints} results in
\begin{equation}\label{car:holonomicConstraintsDerivative}
	\begin{split}
		\dot{x}_2 &= \dot{x}_1 - l\dot{\theta}\sin\theta \\
		\dot{y}_2 &= \dot{y}_1 + l\dot{\theta}\cos\theta,
	\end{split}
\end{equation}
which can be substituted into \eqref{car:KineticEnergy} to find
\begin{equation}
	\begin{split}
		\mathcal{T}^* 
		&= 
		\frac12 m_1(\dot{x}_1^2 + \dot{y}_1^2) + \frac12 J_1\dot{\theta}^2 
		+ 
		\frac12 m_2\left((\dot{x}_1 - l\dot{\theta}\sin\theta)^2 \right.\\
		&\left.\qquad
		+ (\dot{y}_1 + l\dot{\theta}\cos\theta)^2\right) + \frac12 J_2\dot{\phi}_2^2.
	\end{split}
\end{equation}
Taking the configuration to be $ q = (x_1,y_1,\theta,\phi)$, the mass matrix for the car-like system to be
\begin{equation}\label{car:M0}
	M_0( q)
	=
	\begin{bmatrix}
		m_1+m_2 & 0 & -m_2l\sin\theta & 0 \\
		0 & m_1+m_2 & m_2l\cos\theta & 0 \\
		-m_2l\sin\theta & m_2l\cos\theta & m_2l^2+J_1 & 0 \\
		0 & 0 & 0 & J_2
	\end{bmatrix}.
\end{equation}
It is assumed that the system experiences linear viscous damping with dissipation term
\begin{equation}\label{car:D0}
	D_0
	=
	\begin{bmatrix}
		d_u & 0 & 0 & 0 \\
		0 & d_u & 0 & 0 \\
		0 & 0 & d_{\theta} & 0 \\
		0 & 0 & 0 & d_{\phi}
	\end{bmatrix},
\end{equation}
where $d_u, d_{\theta}, d_{\phi} > 0$ are the damping coefficients. $d_u$ is the coefficient for the $x_1$ and $y_1$ directions while $d_{\theta}$ is the damping in the $\theta$ direction and $d_{\phi}$ is the damping in the $\phi$ direction. It is assumed that there is a force input along $\theta$ to the rear wheel and a torque input about $\phi$ on the front wheel, which gives the input mapping matrix
\begin{equation}\label{car:G0}
	G_0( q)
	=
	\begin{bmatrix}
		\cos\theta & 0 \\
		\sin\theta & 0 \\
		0 & 0 \\
		0 & 1
	\end{bmatrix}.
\end{equation}
The system is subject to two nonholonomic constraints that arise due to the non-slip conditions on the wheels:
\begin{equation}\label{car:nonholConstraint}
\begin{split}
	\dot{y}_1\cos{\theta} - \dot{x}_1\sin{\theta} &= 0 \\
	\dot{y}_2\cos(\theta+\phi) - \dot{x}_2\sin(\theta+\phi) &= 0.
\end{split}
\end{equation}
These constraints can be written without $\dot{x}_2$ and $\dot{y}_2$ using the identities \eqref{car:holonomicConstraintsDerivative} and then expressed in the form \eqref{PHS:PfaffianFactor} with the matrix
\begin{equation}\label{car:Gc}
	G_c^\top( q)
	=
	\begin{bmatrix}
		\sin\theta & -\cos\theta & 0 & 0 \\
		\sin(\theta+\phi) & -\cos(\theta+\phi) & -l\cos\phi & 0
	\end{bmatrix}.
\end{equation}
The matrices \eqref{car:M0}, \eqref{car:D0}, \eqref{car:G0} and \eqref{car:Gc} describe the car like system in the form \eqref{PHS:canonicalMomentumPfaffian}.

\subsection{Elimination of Lagrange multipliers}
Now the results of Section \ref{PHS} are applied in order to express the equations of motion of the car-like vehicle without Lagrange multipliers. As per \eqref{PHS:momentumTransformationPartition}, we define the matrix
\begin{equation}
	(G_c^\perp( q))^\top
	=
	\begin{bmatrix}
		1 & 0 \\
		\tan\theta & 0 \\
		\frac1l\sec\theta\tan\phi & 0 \\
		0 & 1
	\end{bmatrix},
\end{equation}
which satisfies $G_c^\perp( q)G_c( q) = 0$. Note that this choice of $G_c$ coincides with the kinematic description of the car-like vehicle studied in \cite{Murray1991a}. Defining
\begin{equation}
	Q( q) 
	= 
	(G_c^\perp( q))^\top
\end{equation}
allows us to express systems dynamics without Lagrange multipliers according to Proposition \ref{PHS:nonholreduc}. The car-like system can now be written in the form \eqref{PHS:ReducMomentum}, where the system matrices are computed according to \eqref{PHS:ReducMomentumMatricies},
	\begin{equation}\label{car:modelq}
		\begin{split}
		Q
		&=
		\begin{bmatrix}
			1 & 0 \\
			\tan\theta & 0 \\
			\frac1l\sec\theta\tan\phi & 0 \\
			0 & 1
		\end{bmatrix} \\
		M
		&=
		\begin{bmatrix}
			a( q) & 0 \\
			0 & J_2
		\end{bmatrix} \\
		D
		&=
		\begin{bmatrix}
			b( q) & 0 \\
			0 & d_{\phi}
		\end{bmatrix} \\
		C
		&=
		\begin{bmatrix}
			0 & c( q)p_1 \\
			-c( q)p_1 & 0
		\end{bmatrix} \\
		G
		&=
		\begin{bmatrix}
			l\cos(\theta_1-\theta_2) & 0 \\
			0 & 1
		\end{bmatrix} \\
		\mathcal{V}( q)
		&=
		0, \\
		\end{split}
	\end{equation}
where
\begin{equation}
	\begin{split}
		a( q)
		&=
		\frac{J_1\sin^2\phi + l^2m_2 + l^2m_1\cos^2\phi}{l^2\cos^2\theta\cos^2\phi} \\
		b( q)
		&=
		\frac{d_\theta\sin^2\phi + d_ul^2 - d_ul^2\sin^2\phi}{l^2\cos^2\theta\cos^2\phi} \\
		c( q) 
		&= 
		\frac{(m_2l^2 + J_1)\sin\phi}{\cos\phi(J_1\sin^2\phi + l^2m_2 + l^2m_1\cos^2\phi)}.
	\end{split}
\end{equation}

\subsection{Coordinate transformation}
The dynamics of the car-like system can be expressed in a different set of generalised coordinates in order to obtain a chained structure. Utilising the transformation proposed in \cite{Murray1991a}, the transformation $f_z$ is defined as
\begin{equation}
	\begin{split}
		z
		&=
		\begin{bmatrix}
			z_1 \\ z_2 \\ z_3 \\ z_4
		\end{bmatrix}
		=
		f_z(q)
		=
		\begin{bmatrix}
			x_1 \\ \frac{1}{l}\sec^3\theta\tan\phi \\ \tan\theta \\ y
		\end{bmatrix},
	\end{split}
\end{equation}
which results in a new pH system of the form \eqref{PHS:ReducMomentumZCoord} with
\begin{equation}\label{system:modelz}
		Q_z
		=
		\begin{bmatrix}
			1 & 0 \\
			\frac{3z_2^2z_3}{z_3^2 + 1} & \frac1l(z_3^2 + 1)^{\frac32}\left[\frac{l^2z_2^2}{(z_3^2 + 1)^3} + 1\right] \\
			z_2 & 0 \\
			z_3 & 0
		\end{bmatrix}.
\end{equation}
The control law can be implemented as a function of $z$ as per \eqref{CtrlLawZ}.

\subsection{Numerical simulation}
\begin{figure}[htbp]
	\centering
	\includegraphics[width=0.5\textwidth]{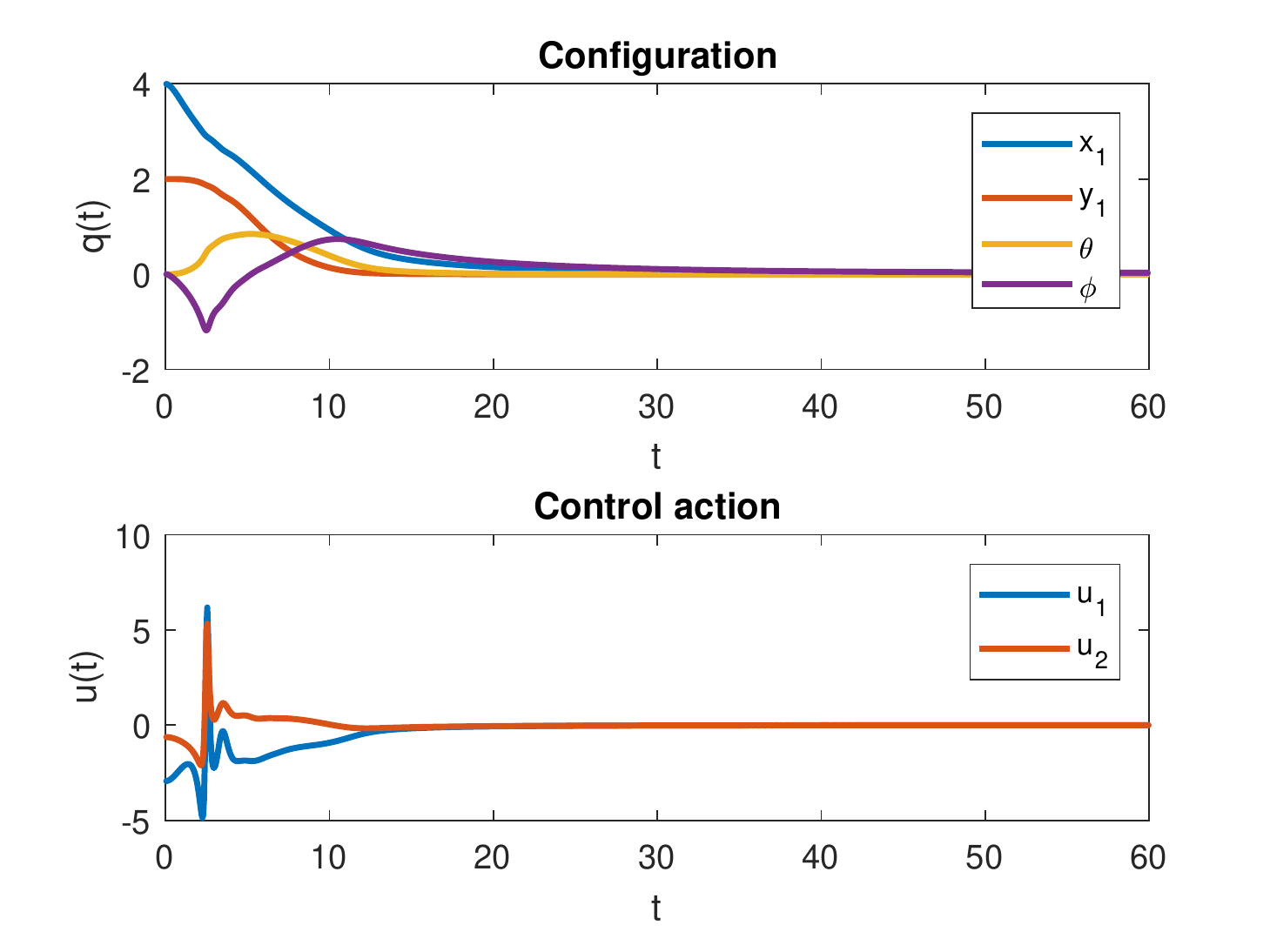}
	\caption{Time history of the configuration variables of the car-like vehicle. Using the discontinuous control law, all states converge to the origin asymptotically.}
	\label{Car:fig1}
\end{figure}

\begin{figure}[htbp]
	\centering
	\includegraphics[width=0.5\textwidth]{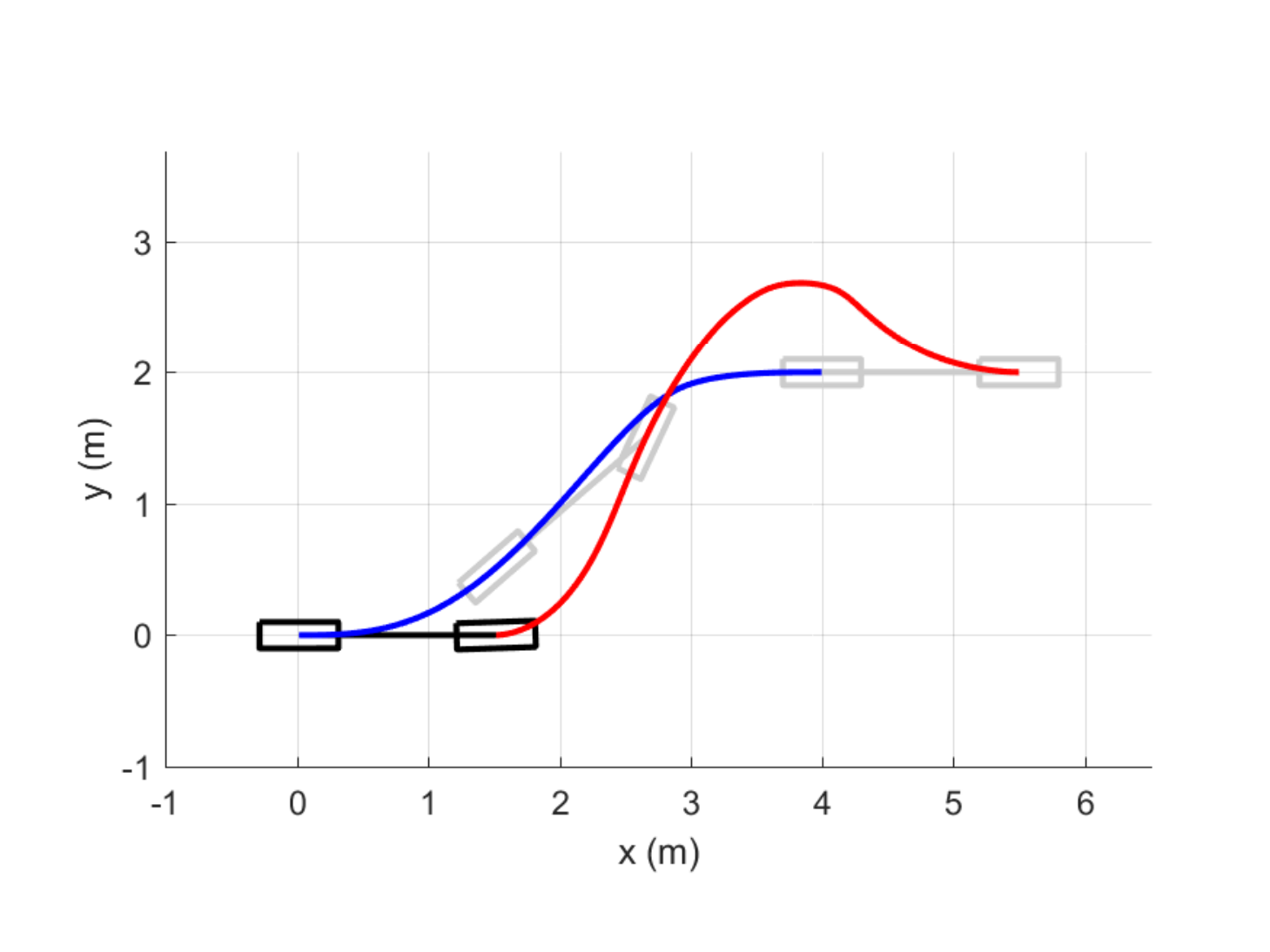}
	\caption{The car-like vehicle moving through the task space. The ghosted images represent a time-lapse of the trajectory at times $t_0 = 0.0s, t_1 = 7.05s, t_2 = 60.0s$. The red and blue lines are the paths of the front and rear wheel projected onto the $x-y$ plane respectively.}
	\label{Car:fig2}
\end{figure}

The car-like vehicle was simulated using the following parameters:
\begin{multicols}{2}
	\noindent
	\begin{equation*}
		\begin{split}
			m_1 &= 0.5 \\
			m_2 &= 2 \\
			J_1 &= 1 \\
			J_2 &= 1 \\
			L &= \operatorname{diag}(1, 10, 0.01, 0.0001)
		\end{split}
	\end{equation*}
	\begin{equation}
		\begin{split}
			l &= 1.5 \\
			d_u &= 4 \\
			d_\theta &= 1 \\
			d_\phi &= 2 \\
			k &= 0.01.
		\end{split}
	\end{equation}
\end{multicols}
The simulation was run for $60$ seconds using the initial conditions $x(0) = 4, y(0) = 2, \theta(0) = 0, \phi(0) = 0$.
Figure \ref{Car:fig1} shows the time history of the states $q(t)$ and control action $u(t)$ and Figure \ref{Car:fig2} is a time-lapse plot of the car-like vehicle travelling from its initial conditions to the origin.

\section{Conclusion}\label{conclusion}
In this paper, a discontinuous control law for set-point regulation of nonholonomic, port-Hamiltonian systems with chained structure to the origin is presented. The control scheme relies on a discontinuous coordinate transformation that reduces the control problem to the stabilisation of a single variable in the transformed space. The proposed control law can be interpreted as potential energy shaping with damping injection and is robust against the inertial and damping properties of the open-loop system. Future work will be concerned with extending the analysis of the closed-loop system to consider the asymptotic behaviour of the momentum variables and control action.


\appendix
\section{Appendix}    
\begin{pf*}{Proof of Lemma \ref{Lemma:QInvert}}
    Assume that $Q_0^\top$ is singular.
    Then, there exists a non-trivial solution $z$ to $Q_0^\top z = 0$.
    Then,
    \begin{align}
        \label{eq:lemAG:a}
        A^\top z &= 0
        \intertext{and}
        \label{eq:lemAG:b}
        G_c^\perp z &= 0
        \text{.}
    \end{align}
    Since $z \neq 0$, then \eqref{eq:lemAG:b} requires that $z = G_c x$ for some $x \neq 0$ as $z$ must be in the range of $G_c$.
    Then \eqref{eq:lemAG:a} becomes $A^\top G_c x = 0$. As $A^\top G_c$ is invertible by assumption, $x$ cannot be non-zero, which is a contradiction. Hence, we conclude that $Q_0$ must be invertible.\qed
\end{pf*}
\vspace{1em}
\begin{pf*}{Proof of Lemma \ref{InvertSn}}
	Consider the matrix
	\begin{equation}
		A_k
		=
		\begin{bmatrix}
			\frac{(n+1)!}{(n-k+2)!} & \frac{(n+2)!}{(n-k+3)!} & \cdots & \frac{(n+k)!}{(n+1)!} \\
			\frac{(n+1)!}{(n-k+3)!} & \frac{(n+2)!}{(n-k+4)!} & \cdots & \frac{(n+k)!}{(n+2)!} \\
			\vdots & \vdots & \ddots & \vdots \\
			\frac{(n+1)!}{(n-1)!} & \frac{(n+2)!}{n!} & \cdots & \frac{(n+k)!}{(n+k-2)!} \\
			\frac{(n+1)!}{n!} & \frac{(n+2)!}{(n+1)!} & \cdots & \frac{(n+k)!}{(n+k-1)!} \\
			1 & 1 & \cdots & 1
		\end{bmatrix}.
	\end{equation}
	As $A_n = S_n\operatorname{diag}\{(n+1)!,(n+2)!,\cdots,2n!\}$, invertibility of $A_n$ is equivalent to invertibility of $S_n$.
	We will show that $A_k$ is invertible by induction. Suppose that $A_{k-1}$ is invertible. Subtracting each column of $A_k$ by the column on its left results in the matrix 
	\begin{equation}
		\tilde A_k
		=
		\begin{bmatrix}
			\frac{(n+1)!}{(n-k+2)!} & \frac{(n+1)!}{(n-k+3)!}(k-1) & \cdots & \frac{(n+k-1)!}{(n+1)!}(k-1) \\
			\frac{(n+1)!}{(n-k+3)!} & \frac{(n+1)!}{(n-k+4)!}(k-2) & \cdots & \frac{(n+k-1)!}{(n+2)!}(k-2) \\
			\vdots & \vdots & \ddots & \vdots \\
			\frac{(n+1)!}{(n-1)!} & \frac{(n+1)!}{n!}(2) & \cdots & \frac{(n+k-1)!}{(n+k-2)!}(2) \\
			n+1 & 1 & \cdots & 1 \\
			1 & 0 & \cdots & 0
		\end{bmatrix}.
	\end{equation}
	Notice that the top right $(k-1)\times(k-1)$ block is $\operatorname{diag}(k-1,k-2,\cdots,2,1)A_{k-1}$ which is invertible by our inductive hypothesis. Thus $\tilde A_k$, and hence $A_k$, is invertible. To complete the proof, we note that $A_1 = [1]$ is trivially invertible. 
\end{pf*}
\vspace{1em}
\begin{pf*}{Proof of Lemma \ref{FuncInequality}}
	By the Schwarz inequality \cite{lieb2001analysis}, any two real valued functions $f(x)$, $g(x)$ satisfy
	\begin{equation}\label{schwartz}
		\left(\int_{x_1}^{x_2}f(x)g(x)dx\right)^2
		\leq
		\int_{x_1}^{x_2}f^2(x)dx
		\int_{x_1}^{x_2}g^2(x)dx.
	\end{equation}
	Taking $g(x) = 1$, \eqref{schwartz} simplifies to
	\begin{equation}
		\begin{split}
			\left(\int_{x_1}^{x_2}f(x)dx\right)^2
			&\leq
			\int_{x_1}^{x_2}f^2(x)dx
			\int_{x_1}^{x_2}1dx \\
			&\leq
			(x_2-x_1)\int_{x_1}^{x_2}f^2(x)dx \\
			\frac{1}{x_2-x_1}\left(\int_{x_1}^{x_2}f(x)dx\right)^2
			&\leq
			\int_{x_1}^{x_2}f^2(x)dx \\
		\end{split}
	\end{equation}
	Taking the negative of this inequality results in
	\begin{equation}
		-\frac{1}{x_2-x_1}\left(\int_{x_1}^{x_2}f(x)dx\right)^2
		\geq
		-\int_{x_1}^{x_2}f^2(x)dx \\
	\end{equation}
	as desired.
\end{pf*}

\end{document}